\documentstyle[aps]{revtex}
\tighten
\newcommand{\xbj}{x_{\scriptscriptstyle B}}

\newcommand{\st}{{\scriptscriptstyle T}}
\newcommand{\ssl}{{\scriptscriptstyle L}}
\newcommand{\bk}{\bbox{k}}
\newcommand{\bkt}{\bbox{k}_{\scriptscriptstyle T}}

\newcommand{\bS}{\bbox{S}}

\newcommand{\bpt}{\bbox{p}_{\scriptscriptstyle T}}
\newcommand{\bppt}{\bbox{P}_{\scriptscriptstyle h\perp}}

\newcommand{\htH}{\tilde H}

\newcommand{\htee}{\tilde e}
\newcommand{\hF}{D}

\newcommand{\hH}{H}
\newcommand{\zh}{z_h}

\newcommand{\ba}{\begin{eqnarray}}
\newcommand{\ea}{\end{eqnarray}}
\newcommand{\be}{\begin{equation}}
\newcommand{\ee}{\end{equation}}



\begin{document}
 
\title{ { \it Sin}$\,\phi$ azimuthal asymmetry in semi-inclusive
  electroproduction on longitudinally polarized nucleon 
  }

\author
{\bf{
    K.A.~Oganessyan$^{a,b}$,
    H.R.~Avakian$^b$
    N.~Bianchi}\\
  {\normalsize \it{LNF-INFN, I-00040, Enrico Fermi 40, Frascati, Italy}}\\ 
  \bf{A.M.~Kotzinian$^{b,c}$}\\
  {\normalsize \it{CERN, CH-1211, Geneva 23, Switzerland}}\\
  }
\maketitle 

\bigskip 

\vspace*{3cm}

\begin{abstract}
We investigate the  {\it sin}$\,\phi$ azimuthal asymmetry in the 
semi-inclusive deep-inelastic lepton scattering off longitudinally polarized
nucleon target arising from the time reversal odd structures. The order 
$1/Q$ contributions of the leading twist and twist-three distribution and 
fragmentation functions to that asymmetry for the certain kinematical 
conditions are numerically estimated.  
\end{abstract}

\vspace*{8cm}
a){ \it{corresponding author, e-mail: kogan@lnf.infn.it} }

b){ \it{On leave of absence from Yerevan Physics 
    Institute, Alikhanian Br.2, \\
    \hspace*{1.0cm} AM-375036 Yerevan, Armenia} }

c){ \it{and JINR, RU-141980 Dubna, Russia} }
\newpage

\section{Introduction}
The combination of transverse momentum and polarization effects in 
the parton distribution and fragmentation functions (DF's and FF's)
results in a rich variety of information on the hadronic 
structure and interactions [1-3]. One specific aspect of the information 
is the appearance of ``time-reversal-odd'' (T-odd) FF's \cite{COL,TM}. 

These T-odd FF's are experimentally accessible via the measurements of 
azimuthal asymmetries in polarized semi-inclusive deep-inelastic
scattering (SIDIS). This kind of 
measurements are currently studied at HERMES experiment at HERA \cite{HM} 
and planed in 
the near future at COMPASS experiment \cite{COMP} at CERN and the proposed 
Electron Laboratory for Europe (ELFE) \cite{ELF}. 
First leading twist numerical estimations for azimuthal asymmetries in
SIDIS have been done in Ref. \cite{KM}. Since the typical values of
$Q^2$ in the above mentioned experiments are not too high it is  
appropriate to evaluate the $1/Q$ effects in the azimuthal asymmetries.

The general expression for azimuthal dependence of polarized SIDIS is
rather complicated \cite{AK}. The complete tree-level description
expression containing contributions from twist-two 
and twist-three DF's and FF's has been given in \cite{TM}.
The purpose of this paper is to investigate the specific $1/Q$ order 
{\it sin}$\,\phi$ azimuthal asymmetry in the SIDIS on a polarized target. 
 
Let $k_1$ ($k_2$) be the initial (final) momentum of the 
incoming (outgoing) charged lepton, $Q^2=-q^2$, $q=k_1-k_2$ the momentum 
of the virtual photon ,
$P_1$ ($P$) the target (observed final-state hadron) momentum. We focus on 
interactions with non-zero outgoing hadron transverse momentum $P_{T}$, 
perpendicular 
to the virtual photon momentum, and denote by $\phi$ the 
azimuthal angle between $P_{T}$ and $k_{1T}$ around the virtual photon 
direction (see Fig.1). 

There are three different
types of contributions to {\it sin}$\,\phi$ asymmetry with the combinations 
of different leading and subleading DF's and FF's.  
 
\begin{itemize}
\item
  The first, often referred to as the fifth structure function,
  contains the twist-three DF $\tilde{e}$ and the time-reversal-odd FF 
  $ H_1^\perp $ and depends on the lepton helicity $ \lambda_e $ \cite{LM}.
  This target spin independent term has been estimated 
  in the Ref.\cite{KOG1} and shown to be negligible. For this reason
  we will not discuss it here.
\item
  The second depends on the longitudinal (with respect to
  virtual photon momentum) component of the target 
  polarization and contains twist-two and
  twist-three DF's and FF's \cite{TM}. 
\item
  Finally, the  
  third containing only twist-two DF and FF is proportional to 
  $\sin(\phi+\phi_S)$ ($\phi_S$ is the azimuthal angle of the 
  target spin vector) and to the transverse (with respect to the virtual 
  photon momentum) target polarization, is known as the ``Collins effect'' 
  \cite{COL}.  In case of target longitudinal polarization it becomes 
  proportional to the transverse component of the target polarization:    
\be
\label{ST}
  S_{\st\,1}=S_{lab} \sin{\theta}_{\gamma}=S_{lab} 
\sqrt{\frac{4M^2x^2_B}{Q^2+4M^2x^2_B}
    (1-y-{M^2x^2_By^2 \over Q^2})},
\ee
  where $M$ is a nucleon mass, $\theta_{\gamma}$ is a virtual photon 
  emission angle and $S_{lab}$ is target polarization (parallel to the 
  incoming lepton momentum). In this case the azimuthal angle ($\phi_S$) 
  of the transverse spin ($S_{\st\,1}$) in the virtual photon frame takes 
  the values $0$ and $\pi$ (see Fig.2) and leads to 
  {\it sin}$\,\phi$ azimuthal asymmetry as well.   
\end{itemize}

\section{The polarized semi-inclusive cross section}

The cross section for one-particle inclusive deep inelastic 
scattering is given by 
\be
\frac{d\sigma^{\ell+N\rightarrow \ell^\prime+h+X}}
{dxdy d\zh d^2P_{h\perp}}=\frac{\pi\alpha^2 y}{2Q^4 \zh} L_{\mu\nu} 2M 
{\cal W}^{\mu\nu} \label{CS},      
\ee
where the kinematical (scale) variables are defined as: 
$$
\xbj = {Q^2 \over {2(P_1q)}}, \quad y={(P_1q) \over (P_1k_1)}, \quad 
\zh={(P_1P) \over (P_1q)}.  
$$ 
The quantity $L_{\mu\nu}$ is the well-known lepton tensor. The full 
expression for the symmetric and antisymmetric parts of the hadronic tensor   
${\cal W}^{\mu\nu}$ at leading $1/Q$ order are given by Eqs. (77), (78) of 
Ref. \cite{TM}. In order 
to investigate the $\sin\phi$ azimuthal asymmetry we keep only the terms 
giving the $\phi$-independent and proportional to $\sin\phi$ 
contributions in the cross section\footnote{to get rid of ambiguities,  
we will use the same notations as in Refs. \cite{TM,BJM}. }:  

\ba
2M\,{\cal W}^{\mu\nu} & = &
2 \zh \int d^2 \bkt\, d^2 \bpt\,
\delta^2(\bpt -{\bppt \over z_h}-\bkt)\nonumber\\ & &
\mbox{}\ \ \times \Biggl\{
-g_\perp^{\mu \nu} f_1 \hF_1
+i\,\epsilon_\perp^{\mu \nu}\,g_{1s} \hF_1
\nonumber\\ & & \quad
+i\frac{2\,\hat t^{\,[ \mu}k_\perp^{\nu ]}}{Q} \Biggl[
-\frac{M}{M_h}\,\xbj\,e \hH_1^\perp
+\frac{m}{M_h}\,f_1 \hH_1^\perp \Biggl]
\nonumber \\ & & \quad
-\frac{\left( k_\perp^{\{\mu} \epsilon_\perp^{\nu \} \rho} S_{\perp\rho}
+ S_\perp^{\{\mu} \epsilon_\perp^{\nu \} \rho} k_{\perp\rho}\right)}{2M_h}
\,h_{1T} \hH_1^\perp
\nonumber \\ & & \quad
+\frac{2\,\hat t^{\,\{\mu }\epsilon_\perp^{\nu\}\rho}k_{\perp \rho}}{Q} \Biggl[
\frac{M}{M_h}\,\xbj \,h_s \hH_1^\perp
- \frac{m}{M_h}\,g_{1s} \hH_1^\perp
\Biggr] \nonumber \\ & & \quad
+ \frac{2\,\hat t^{\,\{\mu }\epsilon_\perp^{\nu\}\rho} p_{\perp\rho}}{Q} \Biggl[
\frac{M_h}{M}\,h_{1s}^\perp \frac{\hH}{\zh}
+ \frac{\bk_\perp^2}{M M_h}\,h_{1s}^\perp \hH_1^\perp
\Biggr] \Biggr\}
\label{WMN}
\ea
where $\{\mu\,\nu\}$ indicates symmetrization of indices and $[\mu\,\nu]$ 
indicates antisymmetrization.  
In the above expression we have used the 
shorthand notation $h_s$  
\ba
h_s (\xbj,\bpt)= \left[ S_\ssl\,h_\ssl\,(\xbj ,\bpt^2)\,
+ h_\st\,(\xbj ,\bpt^2)\,\frac{(\bpt\cdot\bS_\st)}{M} \right].  
\label{long}
\ea 
and similarly for $h^{\perp}_{1s}$ and $g_{1s}$.   
We 
focus on the distributions in a longitudinally polarized nucleon and  
consequently the terms proportional to ${(p_\st \cdot S_\st)}$ of  
Eq.(\ref{long}) vanish upon the integration of cross section over 
$\vec{\bpt}$ ($p_{\st\,1}$ appears only linearly in the cross section). In 
this respect, we will omit them in further.  

The contraction of leptonic and hadronic tensors leads to the cross section 
with the following terms
\be
\frac{d\sigma^{\ell+N\rightarrow \ell^\prime+h+X}}
{dxdyd\zh d^2P_{h\perp}} = \frac{\pi \alpha^2}{Q^2 y} \sum_q e_q^2 
\sigma_q,
\ee
where

\ba
\sigma_q &=& \int d^2p_\st\,d^2k_\st
\, {\zh}^2 \delta^2(\vec{P}_{h\perp} - z_H (\vec{\bpt} - \vec{\bkt})) 
\nonumber \\ 
&& \mbox{} \times \Biggl\{
2\left[1+(1-y)^2\right]\,f^q_1(\xbj,p_\st^2)D^q_1(\zh,\zh^2k_\st^2)
\nonumber \\ 
&& \mbox{} 
+2\lambda_e S_\ssl \,y(2-y)
\,g^q_1(\xbj,p_\st^2)D^q_1(\zh,\zh^2k_\st^2)
\nonumber \\ 
&& \mbox{}
+ 4\lambda_e \,y\sqrt{1-y} \,{1 \over Q} \xbj\, {2M \over M_h} k_{\st\,2} 
\,\htee^q(\xbj,p_\st^2)H^{\perp q}_1(\zh,\zh^2k_\st^2)
\nonumber \\ 
&& \mbox{} - 4(1-y)
\,\frac{ k_{\st\,2} S_{\st\,1}}{M_h} 
\,h^q_{1\st}(\xbj,p_\st^2) H_1^{\perp q}(\zh,\zh^2 \bkt^2)
\nonumber \\ 
&& \mbox{} + {S_\ssl \over Q}\,8(2-y)\sqrt{1-y} 
\,\Biggl [ {M\over M_h}  k_{\st\,2} \Biggl ( \xbj h^q_L(\xbj,p_\st^2)-
{m \over M} g^q_1(\xbj,p_\st^2) \Biggr) H_1^{\perp q}(\zh,\zh^2 k_\st^2)
\nonumber \\ 
&& \mbox{}  
+ {M_h \over M}\, p_{\st\,2} h_{1L}^{\perp q}(\xbj,p_\st^2) 
{\htH^q (\zh,\zh^2 k_\st^2) 
\over \zh} \Biggr ] 
\Biggr\} \label{SCS},  
\ea
where by $k_{\st\,2}$ ($p_{\st\,2}$) we denote the $y$ component of the 
final (initial) parton transverse momentum vectors, and by 
$S_{\st\,1}$-the transverse component of the target polarization 
(see Eq.(\ref{ST})). 

\section{$\bppt$-integrated weighted cross section}
 
Let us consider the differential cross section for one quark flavour, 
$\sigma_q$ , integrated with different weights depending on the final
hadron transverse momenta 
$w_i(P_{h\perp})$:
\be
I_i=\int d^2P_{h\perp}\, w_i(P_{h\perp})\,\sigma_q. 
\ee

Taking into account that  
$I_i = \int d^2k_\st\, d^2p_\st 
\,w_i\left(z(p_\st-k_\st)\right)\{...\}$
and that the odd powers of $k_{\st\,1}$, $p_{\st\,1}$, $k_{\st\,2}$, 
$p_{\st\,2}$ give zero contribution to $I_i$, 
we get cross sections involving
the transverse momentum-integrated distribution and fragmentation
functions. 

\begin{enumerate}

\item
$w_1(P_{h\perp})=1$.

\be
\label{WI1}
I_1=2[1+(1-y)^2]\,f_1(x) D_1(\zh)+2\lambda_e S_L\,y(2-y)\,g_1(x)D_1(\zh),
\ee 

\item
$w_2(P_{h\perp})=-\bppt^2 /{M M_h \zh}=
\vert\bppt\vert\sin\phi/{MM_h\zh}$. 

The surviving terms upon integration are
\ba
\label{WI2}
I_2 = I_{2L} + I_{2T}  \hspace*{8cm} 
\nonumber \\
= {S_\ssl \over Q}\,8(2-y)\sqrt{1-y} 
\,\Biggl [ \Biggl ( \xbj h_L(\xbj)-
{m \over M} g_1(\xbj) \Biggr) H_1^{\perp (1)}(\zh)  
- h_{1L}^{\perp (1)}(\xbj) {\htH (\zh) 
\over \zh} \Biggr ] 
\nonumber \\
+ {{\vert S_{\st\,1}\vert} \over M}\,4(1-y)\,h_1(\xbj) H_1^{\perp (1)}(\zh), 
 \hspace*{4cm} 
\label{RRR}
\ea 
where $I_{2L}$ corresponds to the higher twist effects and $I_{2T}$ is   
representing the leading twist Collins effect. Note, that the 
$S_{\st\,1}$ itself is $\sim 1/Q$ for longitudinally polarized target, 
thus the two contributions are of the same order.

\end{enumerate}

Here we have straightforwardly $\bpt$-($\bkt$)-integrated 
distribution (fragmentation) functions and $\bpt^2/2M^2$- ($\bkt^2/2M^2_h$)-
weighted distribution (fragmentation) functions indicated with 
superscript $(1)$ \cite{TM,MD}:   
\ba
h_{1 \ssl}^{\perp \, (1)}(\xbj)=\int d^2p_\st\,{\left(\frac{p_\st^2}{2M^2}
\right)}\, h_{1 \ssl}^{\perp}(\xbj,p_\st^2), \nonumber \\
H_1^{\perp (1)}(\zh)=\zh^2 \int d^2k_\st\,{\left(\frac{k_\st^2}{2M^2_h}
\right)}\, H_1^{\perp}(\zh,\zh^2k_\st^2) \label{DF}. 
\ea

\subsection{Gaussian parametrizations of DF's and FF's}

The $\bppt$-integration of cross section can also be performed analytically 
if one supposes that the transverse momentum dependence in the DF's 
and FF's be written in factorized exponential form:  

\be
\label{gauss}
 f(\vec{\bpt}) = {1 \over a^2\pi} e^{-{\bpt^2/a^2}}, \qquad 
 d(\vec{\bkt})={1 \over b^2\pi} e^{-{\zh^2\bkt^2}/b^2}.   
\ee

The explicit expression\footnote{We have omitted the 
   charge-square weighted sum over quark flavors $\sum_i{Q_i^2}$.} 
 for the hadronic cross sections up to subleading order $1/Q$ 
(for more details see Appendix D of Ref.\cite{TM}) 
 looks as\footnote {Note that this  expression is identical with
   results in \cite{TM} where the authors have used the following
   relation among DF's  
   $$
   h_L(x_B,k^2_T) = {m \over M} {g_{1L}(x_B,k^2_T) \over x_B}-
   {k^2_T \over M^2} {h^{\perp}_{1L}(x_B,k^2_T) \over x_B} + 
   \tilde{h}_L (x_B,k^2_T).           
   $$
   } 
 
\ba
   \int d{\sigma} d^2P_T &=& C(P_C)
   (1+(1-y)^2)f_1(\xbj) D_1(\zh)+\lambda_e S_\ssl y(1-{y \over 2})g_1(\xbj) 
   D_1(\zh), \label{FD1} \\
   \int d{\sigma} \sin \phi d\phi &=& C(P_T) \biggl \{ S_L
2(2-y) \sqrt{1-y}{\bppt \over {\zh\, Q}}
\biggl [ R_1 \xbj h_\ssl (\xbj) H_1^{\perp}(\zh) \nonumber \\
& - & R_2 g_1(\xbj) H_1^{\perp}(\zh) - 
R_3 h_{1L}^{\perp} (\xbj) {\htH (\zh) \over \zh}\biggr ]  \nonumber \\
& + & S_{\st\,1} (1-y) {\bppt \over \zh} R_4  h_1(\xbj) H_1^{\perp}(\zh) 
\biggr \}\, \label{GHT}.   
\ea
Here 
$$
C(P)={4\pi^2 \alpha^2 \over Q^2 y} \exp(-\frac{P^2}{B})\, , \quad
B=b^2+a^2\zh^2
$$
and 
$$
R_1 = {Mb^2 \over M_h B^2}\, , \quad 
R_2 = {mb^2 \over M_h B^2}\, , \quad 
R_3 = {M_ha^2\zh^2 \over M B^2}\, , \quad
R_4 = {b^2 \over M_h B^2}.
$$
where $M_h$ is the final hadron mass and $m$ is a current quark mass, and 
$S_{\st\,1}$ is the transverse component of the target polarization defined by 
Eq.(\ref{ST}).   

\section{Leading and subleading DF's and FF's} 

In considered SIDIS cross section together with the well-known chiral even 
twist-two DF's and FF's $f_1(\xbj)$ and $D_1(\zh)$ also enter the combinations 
of different leading and subleading DF's and FF's. Their physical 
interpretation has been discussed in Refs.\cite{COL,KM,MD}.

In our numerical calculations we will use the twist-two 
$h_{1L}^{\perp}(\xbj)$, $h_{1L}^{\perp (1)}(\xbj)$, $h_1(\xbj)$, 
$g_1(\xbj)$ and 
twist-three $h_L(\xbj)$ distribution functions obtained within the 
framework of a diquark spectator model of Jacob, Mulders and Rodrigues 
\cite{JMR} (JMR model).

To obtain the time-odd twist-2 ($H_1^{\perp}(\zh)$) and the interaction 
dependent part of the twist-three $H(\zh)$  
($\htH (\zh)$) fragmentation functions we take from the JMR model the 
$\bkt$ depending twist-two fragmentation function 
$D_1^{\pi} (\zh,\zh\vec{\bkt})$:  
\begin{eqnarray}
  \label{D1}
  D_1^{\pi} (\zh,\zh\vec{\bkt}) = N^2_{\pi} \frac{(1-\zh)^{2\alpha-1}}
  {z^{2\alpha}_H} \frac{(m_q+M/\zh)^2+\vert \bkt^2 \vert}
{(\vert \bkt^2 \vert + \lambda^2(1/\zh))^{2\alpha}}, 
\end{eqnarray}
where $\lambda^2 = \Lambda^2(1-1/\zh)+m^2_q/\zh-(1-1/\zh)M^2_h/\zh$, 
$m_q$ is the constituent quark mass, $\Lambda$ 
and $\alpha$ are model parameters. $N^2_{\pi}$ is the normalization factor 
which drops from the ratios of polarized to unpolarized FF's defining the 
$sin\phi$ asymmetry. 

Then we combine that function with the "guess" of Collins \cite{COL}, e.g  
\begin{eqnarray}
  \label{C1}
  H_1^{\perp}(\zh, \zh^2 \bkt^2) = \frac{M_C M_h}{M^2_C + 
    \vert \bkt^2 \vert} D_1(\zh, \zh^2 \bkt^2), 
\end{eqnarray}
where $M_C \simeq 0.3 \div 1.0 $ GeV is a typical hadronic mass. This 
parameterization exhibits the leading twist asymmetry when $\bkt = 
O(M)$. In our numerical calculations we use $M_C=2M_{\pi}$. Using the relation 
\cite{BM}
$$
H(\zh) = \zh^3 {d \over d\zh} {H^{\perp(1)}_1(\zh) \over \zh}
$$    
and Eq.(C.33) of \cite{TM}, one can derive the following expression for the 
interaction dependent part
\begin{equation}
  \label{RRA}
  {\htH (\zh) \over \zh} = {d \over d\zh} [\zh H^{\perp(1)}_1(\zh)],
\end{equation}
where the function with upper index $(1)$ indicate 
$\bkt^2/2M^2_h$-weighted function defined as in Eq.(\ref{DF}). 

The plots $a$, and $b$ of the Fig.3 show $h_L(\xbj)$, $h_1(\xbj)$, and 
$h_{1L}^{\perp} (\xbj)$, $h_{1L}^{\perp(1)}(\xbj)$ distribution functions 
obtained within JMR model. On the plots $c,d$ of the Fig.3 the 
$H^{\perp}_{1}(\zh)/D_1(\zh)$ and 
$H^{\perp(1)}_{1}(\zh)/D_1(\zh)$ using the JMR model for 
$\Lambda = 0.4 GeV$, $m_q = 0.36 GeV$ and $\alpha=1.05$ are presented (labeled 
by $1$). The predictions for the same quantities \cite{KM} using the 
Binnewies, Kniehl, Kramer, (BKK) 
parametrization \cite{BKK} $D_1(\zh)$ for the quark fragmentation 
function, are labeled by $2$. In that approach Collins ``guess'' combined with 
the BKK $D_1(\zh)$ of a Gaussian $\bkt$ dependence.    

\section{Numerical results}

We start with the consideration of specific spin-dependent azimuthal asymmetry 
in polarized $\bppt$-integrated semi-inclusive charged pion  
leptoproduction. Note that in our calculations we make an approximation in 
which we do not take 
into account the sea-quark contributions, as well as unfavourite 
fragmentation functions 
($D_1^{d\to{\pi^{+}}}(\zh)=D_1^{u\to{\pi^{-}}}(\zh)$) have been set to 
zero (consequently, unfavourite time-odd ones as well). In view of these our 
numerical results for the $\sin \phi$ asymmetry should be regarded as a 
order-of magnitude estimates. 
 
Consider the weighted $\sin \phi$ asymmetry defined as
\ba
\langle \frac{\vert P_{h\perp}\vert}{MM_h\zh}
\sin \phi_h \rangle \equiv 
\frac{\int d^2P_{h\perp} \frac{\vert P_{h\perp}\vert}{MM_h\zh}
\sin \phi_h \frac{d\sigma}
{dxdyd\zh d^2P_{h\perp}}}
{\int d^2P_{h\perp} \frac{d\sigma}
{dxdyd\zh d^2P_{h\perp}}} = \frac{I_2}{I_1}
\label{AS},
\ea
where $I_1$, and $I_2$ are given by Eq.(\ref{WI1}) and Eq.(\ref{WI2}). Note, 
that this expression define the target longitudinal spin azimuthal asymmetry 
for unpolarized lepton beam ($\lambda_e = 0$) as well. 

For numerical estimation of asymmetry, defined by Eq.(\ref{AS}), we use 
the following two sets of DFs and FFs: 
 
\begin{itemize}
\item
  {\it set} {\bf A} - the $f_1(\xbj)$, $h_{1L}^{\perp(1)}(\xbj), h_1(\xbj), 
  g_1(\xbj)$, $h_L(x_B)$ distribution and $D_1(\zh)$ fragmentation functions 
  obtained within the JMR model \cite{JMR}. The time-reversal odd 
  fragmentation functions defined by Eqs. (\ref{D1}) - (\ref{RRA}) and 
  Eq.(\ref{DF}). 

\item
  {\it set} {\bf B} - the Gl$\ddot{u}$ck, Reya, Vogt, (GRV) parton 
  distribution functions \cite{GRV} for $f_1(\xbj)$ and the  
  $h_{1L}^{\perp(1)}(\xbj), 
  h_1(\xbj), g_1(\xbj)$, $h_L(x_B)$ distribution functions obtained within 
  the JMR model \cite{JMR}. BKK parametrization of the $D_1(\zh)$ and the 
  time-reversal odd fragmentation functions obtained by combining 
  Collins ``guess'' with the BKK $D_1(\zh)$ of a Gaussian $\bkt$ dependence.  
   
\end{itemize} 

In order to make an average over the range 
of $Q^2$, we use the relation 
$Q^2=2ME_lx_Hy$, where $M$ is the proton mass. After integrating over the 
$\xbj,y,\zh$ at HERMES experiment kinematical ranges, e.g., $E_l=27.5 GeV$, 
$Q^2 > 1 GeV^2$, $0.1 < y < 0.85$, $0.02 <x_H< 0.4$ and $0.1 < \zh < 1$, we 
get\footnote{Note, that in our numerical calculations we neglect the 
contribution of the term $\sim g_1(\xbj)H_1^{\perp(1)}(\zh)$  
suppressed by the factor $m/M$, where $m$ is the current quark mass and 
$M$ the proton mass.}

\ba 
\langle \frac{\vert P_{h\perp}\vert}{MM_h\zh}
\sin \phi_h \rangle = \left \{ 
\begin{array}{r@{\quad:\quad}l} 
0.15 & A \\ 0.17 & B 
\end{array} 
\right. 
\label{WAS} 
\ea 

It is important to mention that the contribution of $I_{2L}$ is about 
$11\%$ in case of using the {\it set} {\bf A} 
and about $12\%$ when using the {\it set} {\bf B}. 

In the same approach using the expression for $I_{2T}$ defined in 
Eq.(\ref{WI2}) we estimate also the magnitude of the weighted asymmetry in 
the cross section (Eq.\ref{AS} with $S_\ssl=0$) related directly to 
so-called ``Collins effect'' for transversely polarized target. 
For {\it set} {\bf A} we get it $0.17$ and for {\it set} {\bf B} it 
is $0.22$. 

Now let us define the quantity $\langle \sin\phi \rangle$ as 
\ba
\langle \sin\phi \rangle = \frac{\int d\sigma 
 \sin \phi d\phi }
{\int d\sigma d\phi },
\ea 
where $d\sigma$ is a SIDIS cross section and 
the integrations are over $P_{T}$ (with the lower limit of
the observed hadrons transverse momentum cutoff equal to $P_C$),  
$\phi$, $\xbj$, $y$ and $\zh$. 

Consider how $\langle \sin\phi \rangle$ as defined above  
with $P_{T}$ cutoff $P_C$ (only hadrons with transverse 
momenta above the cutoff will be included), behaves numerically. We use 
the Gaussian transverse momentum parametrizations of DF's and FF's defined 
by Eqs.(\ref{gauss}) of the same leading and subleading DF's and FF's 
mentioned above.  

Our numerical results at HERMES kinematics mentioned 
above are presented in Fig.4. We take $a=0.5 GeV, b=0.7 GeV$, which 
correspond to an average intrinsic transverse momenta of $\langle k_T 
\rangle = 0.44 GeV, \langle p_{T} \rangle = 0.62 GeV$. This choice of the 
average transverse momenta used to estimate the smearing effects are 
conditioned by 
the good agreement with existing data on the azimuthal structure of the 
hadronic final state in unpolarized deep inelastic $lp$ scattering 
\cite{EMC,E665}. The curves labeled by $1$ and $2$ on the Fig.4 are 
corresponded to the {\it set} {\bf A} and  {\it set} {\bf B}, respectively. 

From Fig.4 one can see that the 
magnitude of such single asymmetry is about a few percents (while at the same 
kinematical range the asymmetry of the semi-inclusive pion 
production in the deep-inelastic scattering of a polarized lepton beam off  
an unpolarized nucleon target shows up also as a $\langle \sin\phi \rangle$, 
is around half-percent \cite{KOG1}). We point out that the big asymmetry 
predicted using the {\it set} {\bf A} is mainly conditioned by the strong 
dependence (especially at small $\bkt$) of the FFs, based on the JMR model, 
on $\bkt$, whereas the $\bkt^2$ factor in the weighted T-odd FFs washed out 
that dependence.     

It is important to emphasize that the 
azimuthal structure in the quark transverse momenta plays the key role in 
this asymmetry.  At HERMES kinematical conditions we deal 
with the low transverse momentum range, where the effects of intrinsic 
transverse momentum are dominated. In this respect we assume the 
estimations within the Gaussian transverse parametrizations of DF's and FF's 
are reasonable. Another argument is that they are in a good agreement with the 
numerical results of the $\bppt$-integrated weighted $\sin \phi$ asymmetry, 
which are valid for {\it any} transverse momentum dependence of DF's and 
FF's.    
      
\section{Summary}
We have investigated the specific $1/Q$ order 
{\it sin}$\,\phi$ azimuthal asymmetry of single inclusive charged pion 
production in the deep-inelastic lepton scattering off longitudinally 
polarized nucleon target related to the time reversal odd structure, 
arising from nonperturbative final-state interactions. The contributions 
of both leading twist and twist-three effects to that asymmetry are taken 
into account. 

The order $1/Q$ $\bppt$-integrated weighted azimuthal asymmetry in terms 
of the leading twist and twist-three distribution and fragmentation 
functions is numerically estimated at HERMES kinematical conditions.   

We have also analyzed the dependence of the azimuthal asymmetry parameter 
$\langle \sin\phi \rangle$ on the transverse momentum cutoff $P_C$ in 
kinematical ranges at HERMES. 

The measurement of such a single spin asymmetry can allow to determine the 
naively-time-odd quark fragmentation functions, which appears due to the 
non-applicability of time-reversal invariance for the hadronization of a 
quark. It is important, however, to have good particle identification and 
sufficient azimuthal resolution in the forward direction. 

\section{Acknowledgments}
The authors would like to thank E. DeSanctis for helpful comments. One of the 
authors (K.O) wish to thank D. Boer, R. Jakob and P. Mulders for useful 
discussions.  This work of (K.O), (H.A) and (N.B) was in part supported by 
the INTAS contribution (contract number 93-1827) from the European Community.

\newpage

\section{Figure captions}

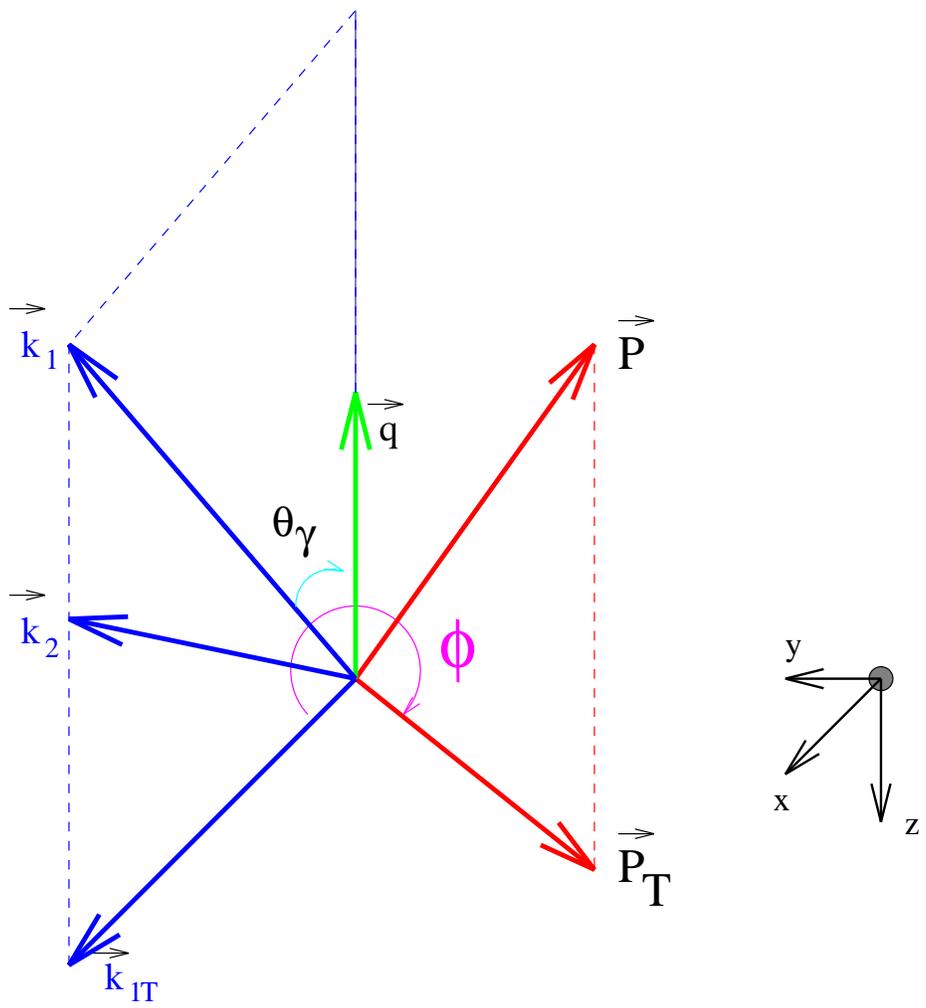
Fig.1. The definition of the final hadron azimuthal angle $\phi$. 

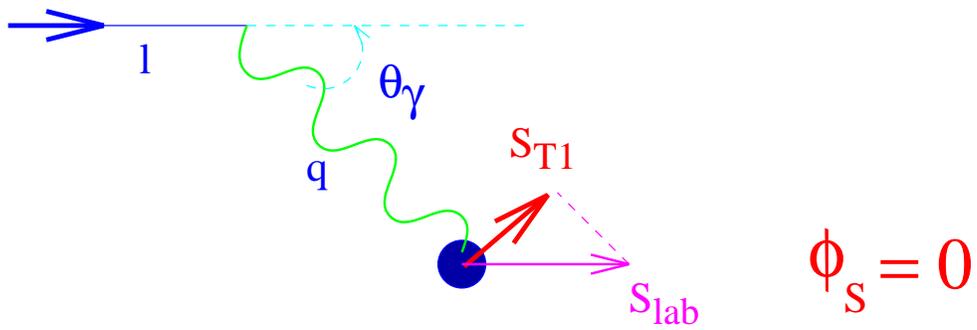
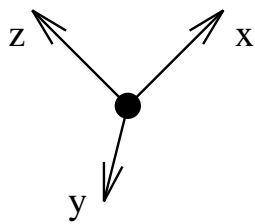
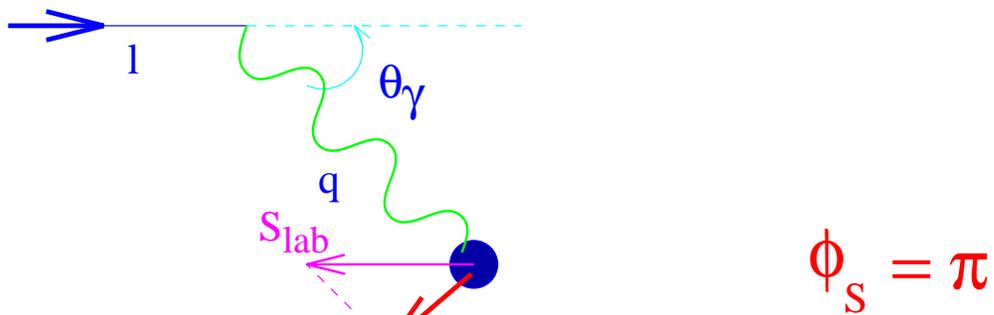
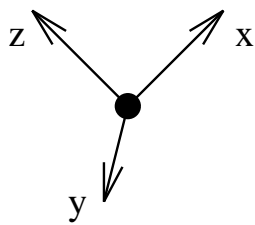
Fig.2. Definition of the transverse spin azimuthal angle in the virtual 
       photons frame.

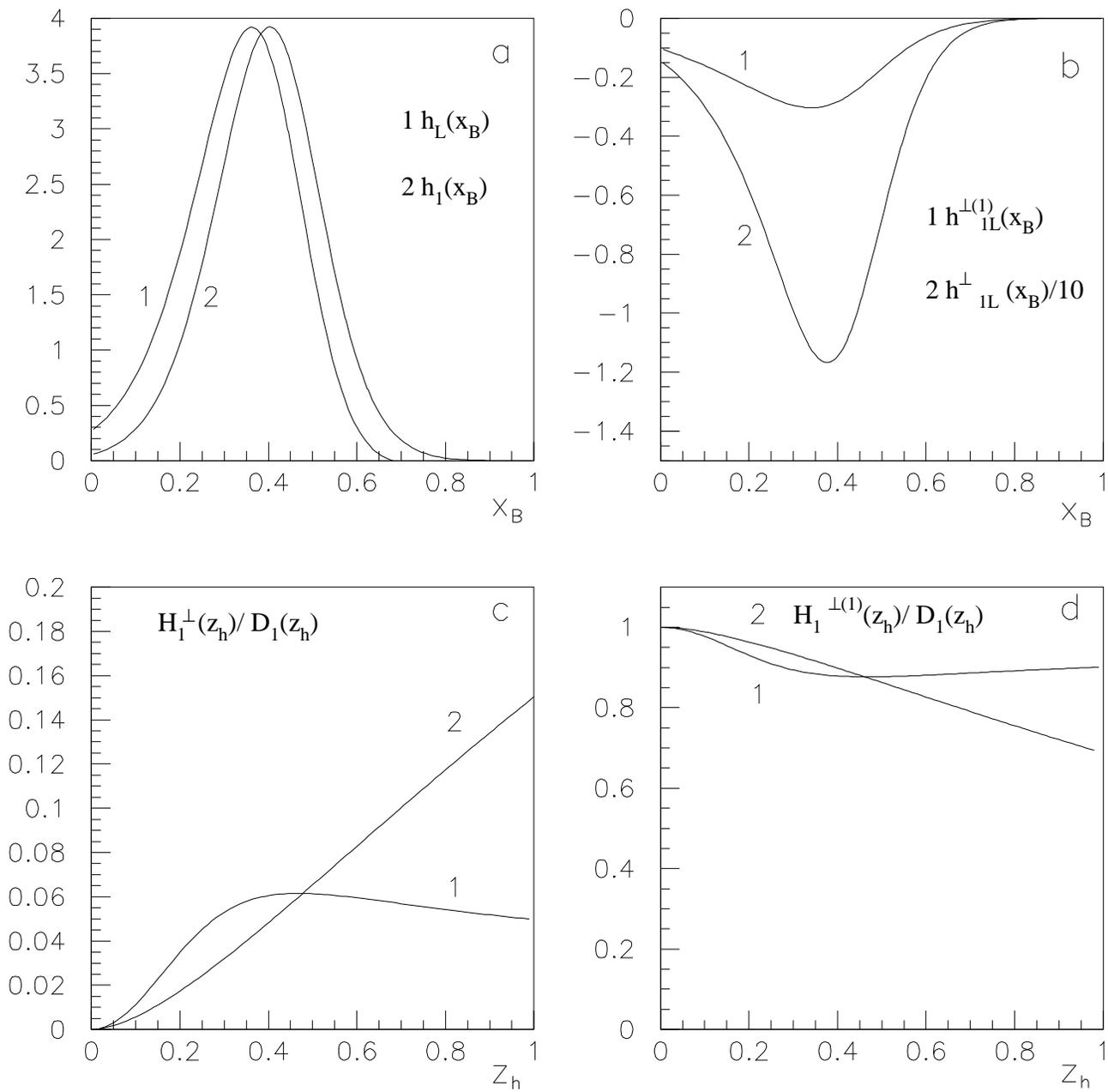
Fig.3. $a$ - $h_L(\xbj)$, $h_1(\xbj)$, $b$ - $h_{1L}^{\perp} (\xbj)$,  
       $h_{1L}^{\perp(1)}(\xbj) 10^{-1}$ distribution functions obtained 
       within JMR model. On $c$ the  
       $H^{\perp}_{1}(\zh)/D_1(\zh)$,  
       and on $d$ the $H^{\perp(1)}_{1}(\zh)/D_1(\zh)$ are ploted. The 
       curves denoted by $1$ and $2$ correspond to ratio of polarized 
       and unpolarized fragmentation functions of the {\it sets} {\bf A} and 
       {\bf B} (see text), respectively.     

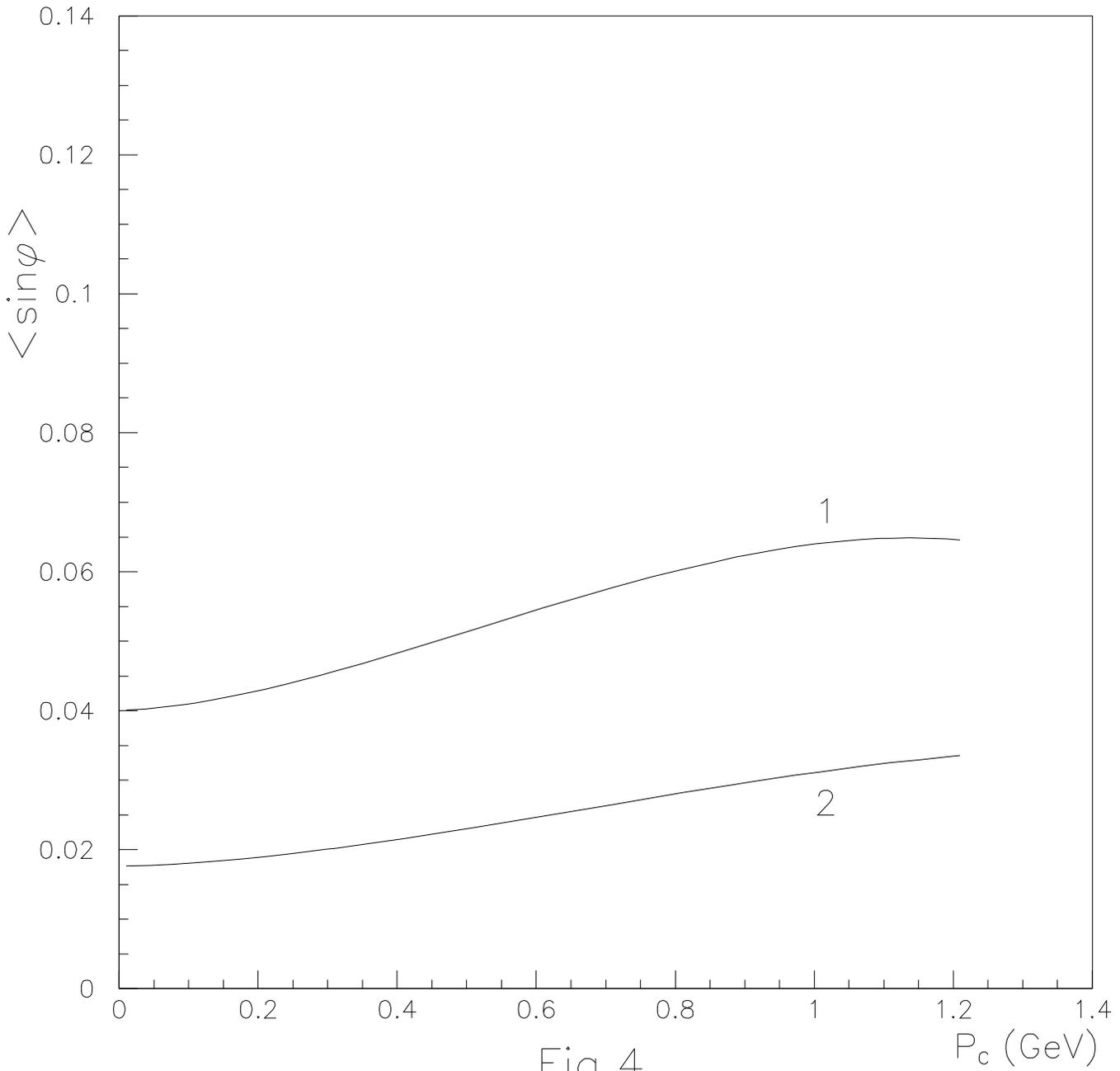
Fig.4. $\langle \sin\phi \rangle$ at HERMES kinematics.  

\end{document}